# Circumventing the Stability Problems of Graphene Nanoribbon Zigzag Edges

James Lawrence[1,2,†], Alejandro Berdonces-Layunta[1,2,†], Shayan Edalatmanesh,[3] Jesús Castro-Esteban,[4] Tao Wang[1,2], Mohammed S. G. Mohammed[1,2], Manuel Vilas-Varela,[4] Pavel Jelinek,[3,*] Diego Peña[4,*], Dimas G. de Oteyza[1,2,5,*]

[1]Donostia International Physics Center; San Sebastián, Spain

[2]Centro de Física de Materiales; San Sebastián, Spain

[3]Institute of Physics, Czech Academy of Sciences; Prague, Czech Republic

[4]Centro Singular de Investigación en Química Biolóxica e Materiais Moleculares (CiQUS) and Departamento de Química Orgánica, Universidade de Santiago de Compostela; Santiago de Compostela, Spain

[5]Ikerbasque, Basque Foundation for Science; Bilbao, Spain

* Corresponding author. Email: jelinekp@fzu.cz; diego.pena@usc.es; d_g_oteyza@ehu.es
† These authors contributed equally to this work

**Abstract:**

Carbon nanostructures with zigzag edges exhibit unique properties with exciting potential applications. Such nanostructures are generally synthesized under vacuum because their zigzag edges are unstable under ambient conditions: a barrier that must be surmounted to achieve their scalable exploitation. Here, we prove the viability of chemical protection/deprotection strategies for this aim, demonstrated on labile chiral graphene nanoribbons (chGNRs). Upon hydrogenation, the chGNRs survive an exposure to air, after which they are easily converted back to their original structure *via* annealing. We also approach the problem from another angle by synthesizing a chemically stable oxidized form of the chGNRs that can be converted to the pristine hydrocarbon form *via* hydrogenation and annealing. These findings may represent an important step toward the integration of zigzag-edged nanostructures in devices.

Graphene nanoribbons (GNRs) are exceptionally versatile materials. Adjusting their length, width, edge structure or heteroatom dopants can have significant effects on their electronic properties, which may vary from being metallic to semiconducting, shifting the band structure up and down in energy, or even endowing the nanoribbons with magnetic properties.[1–3] The possibilities further multiply with heterostructures, formed both out of different GNRs,[4–6] as well as in combination with other functional carbon-based materials.[7–9] As a result, GNRs are extremely promising building blocks for a multitude of applications, including quantum technologies. However, some of the most praised properties of GNRs rely on the presence of zigzag edge segments,[10,11] which, as opposed to armchair edges,[12,13] lack sufficient chemical stability to withstand air exposure even for structures with a dominantly closed-shell character.[14] This severely jeopardizes their potential utilization in actual devices, bringing up the need to conceive alternative strategies for the device implementation processes.

Edge groups have been often utilized with GNRs to make them soluble[15,16] or to tune the ribbon's electronic properties.[17–19] However, although frequently used in solution chemistry with smaller molecules,[20,21] to date their use for protection purposes on GNRs has remained at a theoretical level.[22–24] Besides, it remains to be tested whether such chemical protection strategies can also be applied to on-surface synthesis,[25–27] which is the approach whereby most of the carbon-based nanostructures with peripheral zigzag edges have been synthesized to date.[28,29] Interestingly, such zigzag edges often appear passivated by extra hydrogen atoms,[30,31] causing a rehybridization of the associated carbon atoms into an $sp^3$ configuration and increasing the structure's stability. On the other hand, it has been shown that hydrogenation can also be used for edge modification of nanoribbons; in particular, it has been used to remove peripheral chlorine[32] as well as sulfur atoms,[33] with tweaks to the temperature and hydrogen exposure during different stages of the GNR growth also altering their length and termination.[33]

This raises the question of whether an intentional hydrogenation of otherwise air-sensitive GNRs, like chiral nanoribbons displaying a regular alternation of three zigzag and one armchair unit along their edges ((3,1)-chGNRs),[14] may act as a protective functionalization that could eventually be controllably removed *e.g.* by annealing treatments (**Scheme 1**). In this work, we indeed demonstrate the usage of atomic hydrogen as a means of protecting (3,1)-chGNRs from the oxidizing effects of the atmosphere. A closely related approach further allows converting an air-stable, chemically modified form of the chGNRs with protective ketone side groups, back to pristine nanoribbons. Using the power of bond-resolving scanning tunnelling microscopy (BR-STM), along with scanning tunnelling spectroscopy, we show that both chemically protected forms of the GNRs (hydrogenated and ketone-functionalized) survive exposure to air. The GNRs that follow the deprotection processes are mostly pristine, with their electronic properties remaining intact.

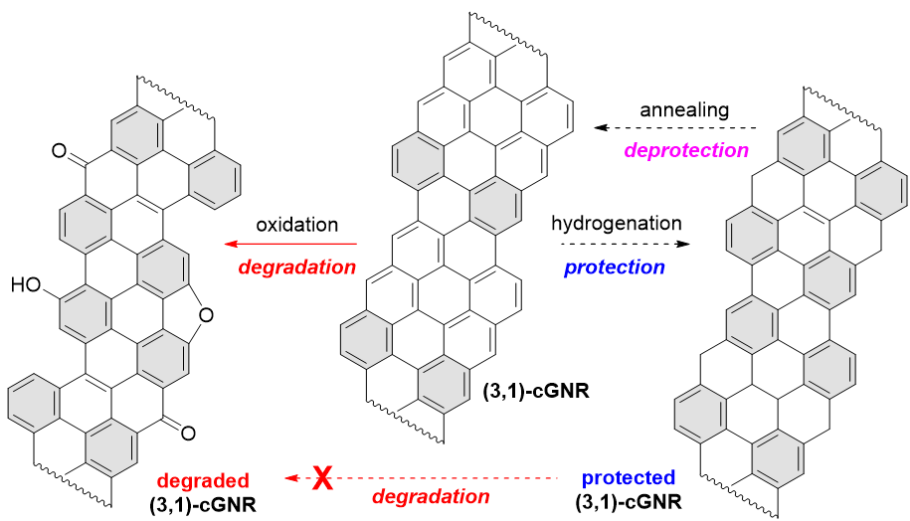

**Scheme 1**. **Representation of the degradation of (3,1)-chGNRs exposed to air and the prospective protection strategy by controlled hydrogenation**. Hydrogenation of the most reactive carbon atoms at the centre of each zigzag segment[14] would confer the ribbons two extra Clar sextets (marked in grey) per unit cell, that are normally associated with larger band gaps and increased stabilities.

**Results**

*Hydrogenation for protection and annealing for deprotection*

The synthesis of chGNRs on a Au(111) surface was performed according to previous protocols,[34,35] *via* annealing 2,2'-dibromo-9,9'-bianthracene (chiral DBBA) precursors up to 350°C. Following this, the resulting nanoribbons were examined *via* BR-STM to ensure that they were pristine. Representative images of the pristine ribbons are shown in **Fig. 1(a)** and **(b)**. In order to protect the edges by hydrogenation, the GNRs were then exposed to a flux of atomic hydrogen at a chamber pressure of $1 \times 10^{-7}$ mbar for 2 minutes (with the sample held at room temperature), created by passing molecular hydrogen through a heated tungsten tube (80 W e-beam heating power, approximately 2800 K). Following this treatment, the GNRs were examined again at 4 K via STM measurements. The result of this GNR protection is shown in **Fig. 1(c)**. Whereas the pristine chGNRs are typically adsorbed separately, after the hydrogen exposure they exhibit a noticeable tendency to aggregate together. Notably, substantial changes in the apparent height contrast within the ribbons are observed, as can be seen best in the inset of **Fig. 1(c)**. **Figure S1** in the supplementary materials shows a comparison between the apparent height profile of pristine and hydrogenated ribbons, with the ribbons approximately doubling their apparent height after hydrogenation. This implies that a significant number of the carbons are converted from $sp^2$ to $sp^3$ upon hydrogen exposure, adopting a non-planar conformation to accommodate the change in favoured bond angles. A comparison to previous reports in the literature, in which some of the edge carbons of chGNRs were shown to be hydrogenated but did not show significant increases in their height profile,[14,30] leads to the conclusion that several internal double bonds of the GNR backbone have also been hydrogenated, leading to sections of 'graphane' nanoribbon that notably differ from the idealized picture in **Scheme 1**. Furthermore, this conversion is not uniform, as can be seen from the topographic images in **Fig. 1(c)** and the aforementioned height profiles.

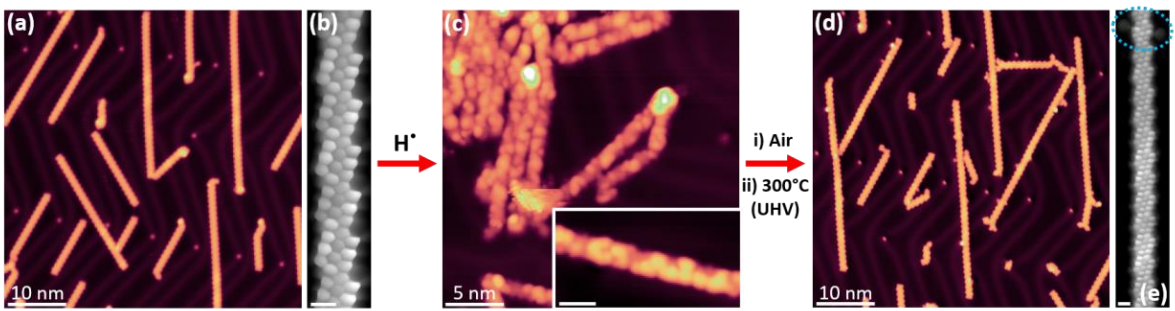

**Figure 1. STM analysis of GNRs along the various steps in the cycle of synthesis, hydrogenation (protection), air-exposure and annealing (deprotection).** (a) Overview STM image of the sample of pristine (3,1)-chGNRs on Au(111) prior to hydrogenation. I = 1 nA, $V_{bias}$ = −0.5 V. (b) BR-STM image of a pristine NR. CO tip, constant height, $V_{bias}$ = 5 mV; scale bar = 500 pm. (c) STM image of a cluster of hydrogenated ribbons after exposure to atomic hydrogen. Inset: zoom of a hydrogenated nanoribbon, showing strong changes in contrast in sections with different levels of hydrogenation. For both images, I = 23 pA, $V_{bias}$ = −2.0 V. Scale bar of zoom = 2 nm. (d) Overview STM image of the hydrogenated nanoribbons after 24 minutes of air exposure followed by annealing to 300°C in UHV. I = 20 pA, $V_{bias}$ = −0.5 V. (e) BR-STM image of a mostly pristine nanoribbon after the same treatment, with one defective/oxidized section with metal-coordinated ketone groups circled. CO tip, constant height, $V_{bias}$ = 5 mV; scale bar = 500 pm.

In order to test the air stability of the hydrogenated nanoribbons, the sample was then taken out of its ultra-high vacuum (UHV) environment and exposed to the air for 24 minutes (see methods). Following this, it was transferred back to UHV (pressure in the low $10^{-10}$ mbar range) and annealed to 300°C for 20 minutes to both dehydrogenate (*i.e.* "deprotect") the GNRs and remove coadsorbed contaminants from the air. The result of this procedure is shown in **Fig. 1(d)** and **(e)**. The vast majority (approx. 90%) of the chGNR units cells are found to be pristine, also clearly retaining their electronic properties,[36] as shown in **fig. S2** with undisturbed conductance maps at the valence and conduction band onsets. A few defects are observed – most commonly, the central ring of the zigzag edge appears larger in BR-STM imaging, with a sharp point at the edge that is oriented towards a feature that presumably corresponds to a gold adatom (**fig. S3**). This is similar to the most commonly observed defect during oxidation experiments with the pristine chGNRs,[14] which has been assigned to a ketone that is coordinated to a gold adatom. An example thereof is circled in blue in **Fig. 1(e)**, with a pair of these ketone defects on the same GNR unit cell. As the number of defects is very low, this result clearly demonstrates the effectiveness of the hydrogenation strategy in protecting the GNRs from the oxidizing effect of the atmosphere; previous experiments observed significant levels of oxidation even at low pressures ($10^{-5}$ mbar) of pure molecular oxygen when instead exposing the pristine chGNRs.[14]

*Pre-oxidized nanoribbons*

To further understand the tendency of chiral GNRs to oxidize at their zigzag edges, we have utilized a different precursor molecule: 2,2'-dibromo-10H,10'H-[9,9'-bianthracenylidene]-10,10'-dione ((*E/Z*)-k-DBBA, see supplementary materials for synthetic details). This molecule is 'pre-oxidized' with ketone groups bound at its central zigzag edge positions (**Fig. 2(a)**), which were previously found to be the most reactive sites.[14] The self-assembly of a mixture of *E* and *Z* isomers of k-DBBA is presented in **fig. S4**. Annealing this precursor mixture on Au(111) to relatively high temperatures (430°C) results in the formation of nanoribbons, henceforth referred to as ketone-chGNRs (k-chGNRs). As shown in **Fig. 2(b)**, these nanoribbons readily pack together into islands. The chemical structure of the k-chGNRs is shown in **Fig. 2(c)**. Two different forms of these nanoribbons are observed: 'normal' k-chGNRs, which are typically found in islands held together by hydrogen bonds; and 'metal-organic' k-chGNRs, found in islands held together by apparent metal-organic bonds. In the case of the latter, the ketone groups are observed pointing towards each other, implying the presence of a shared metal atom, whereas in the former the ketones are oriented towards the hydrogen atoms of an adjacent ribbon. A comparison showing models of both types is shown in **fig. S5**. Although both forms are always present, their ratio varies depending on the exact annealing procedure used during their formation. Annealing straight to a high temperature typically yields mostly metal-organic GNRs, whereas annealing in steps at relatively high temperatures (e.g. 370 °C, 400 °C, 430 °C) yields more 'normal' k-chGNRs. In all experiments, there was a notable issue with anthrone-related molecules (described in further detail in **fig. S6**). These molecules are often found at the termination of the k-chGNRs, and may be the cause of their relatively short average length when compared to pristine chGNRs that are formed from chiral DBBA. We hypothesize that the Z-isomer, which does not form GNRs, may be the cause of this contamination. It may terminate the polymers and fragment when heated on the surface. Future studies with stereoisomerically pure precursors[35] may clarify this issue and improve the length and quality of the nanoribbons.

An example of a BR-STM image of the 'normal' type of k-chGNRs is shown in **Fig. 2(d)**. Each unit cell exhibits two sharp features located at the centre of its zigzag edge, corresponding to the positions of the ketones. These features match with previously observed oxidation defects on pristine chGNRs,[14] as well as those described above after exposing the hydrogenated GNRs to air. Crucially, these pre-oxidized chGNRs are no longer susceptible to oxidation under ambient conditions; after exposing a sample of k-chGNRs to air (24 minutes) and post-annealing (200°C) to remove coadsorbed contaminants (**fig. S7**), the vast majority of the k-chGNR units are unaffected, with only a small minority (10%) displaying any extra defects that could be ascribed to the air exposure. BR-STM images of some of these defects are shown in **fig. S8**. A clear demonstration of the general air stability of the k-chGNRs is the image shown in **Fig. 2(d)**, which was actually recorded after air exposure followed by annealing, and displays no changes from typical BR-STM images recorded prior to the exposure. It may well be the case that with a better preparation of k-chGNRs, the level of defects originating from the air exposure could be further reduced. This is because, after their initial synthesis, one of the more common defects observed is the absence of a ketone group on one or both sides of a unit cell. This presumably renders the unit more susceptible to attack by oxidative species in the air, resulting in a greater number of oxidation defects after exposure. Of course, some of these units most probably go on to form ketone groups anyway, as this was seen to be one of the most common defects observed after the exposure of pristine chGNRs to air.[14]

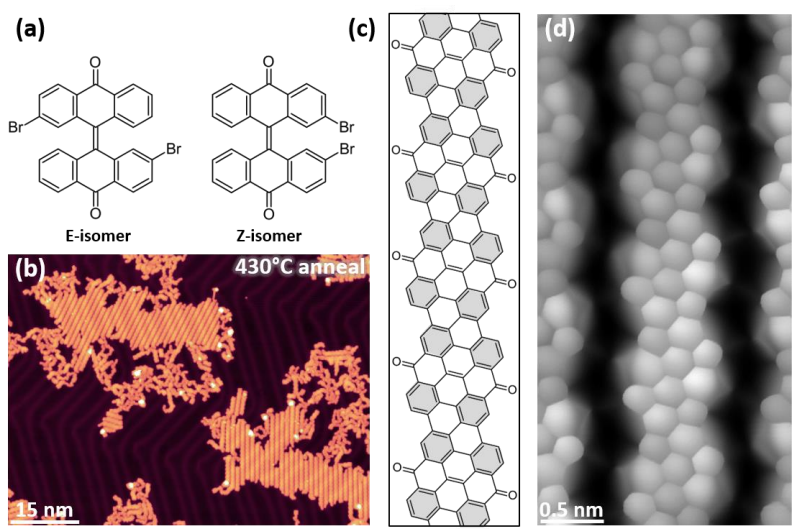

**Figure 2. Reactant and product structure of pre-oxidized (protected) chiral GNRs.** (a) Structures of the two k-DBBA precursor isomers. (b) Overview STM image of self-assembled islands of ketone chiral nanoribbons after annealing the precursors in steps up to 430°C. I = 30 pA, $V_{bias}$ = −0.5 V. (c) Chemical structure of the ketone chiral nanoribbons, with Clar sextets marked by a grey background (d) Example of a BR-STM image of a ketone chiral nanoribbon after air exposure and post-annealing to 200°C in UHV conditions. The sharp features at the edge of each central ring correspond to the ketone groups, and these were observed both before and after the air exposure. CO tip, constant height, $V_{bias}$ = 8 mV.

As the addition of a ketone group significantly changes the conjugation of the chGNRs, it follows that the measured electronic properties are strongly affected. Each unit cell can be represented with four Clar sextets (**Fig. 2c**), as opposed to the two Clar sextets that may be drawn for the (closed shell) representations of the pristine (3,1)-chGNRs (**Scheme 1**), an increase that is normally associated with an enlarged band gap and a concomitantly augmented stability. A 45-point line of dI/dV spectra, recorded along a 'normal' 10-unit k-chGNR, is presented in **Fig. 3(a)**. Overlaid is an example of one of the point spectra from the line, recorded at the position marked with a red cross in **Fig. 3(b)**. The onsets of both the valence and conduction band are indicated with dashed lines. Besides conferring the ribbons an n-type semiconducting character, the ketone functionalization endows the ribbons with a much larger band gap (approximately 2.1 eV) than that of the pristine chGNRs, which tends to approximately 0.65 eV.[36] STS measurements of k-chGNRs of different lengths show that their band gap very quickly saturates to around 2.05 – 2.10 eV after reaching 4 or 5 units in length (**fig. S9**). Indeed, the observed HOMO-LUMO gap of a dimer was found to be only 2.5 eV. This notably wider band gap supports the assertion of the k-chGNRs' increased stability with respect to the pristine chGNRs. A BR-STM image of the 10-mer, along with four constant height dI/dV maps (CO tip), are presented in **Fig. 3(b)-(f)**. These are recorded at the onsets of the valence and conduction bands, and at two conductance maxima that are found deeper into the occupied and unoccupied states.

To improve our understanding of the ribbon's electronic properties we have performed complementary DFT calculations. **Fig. 3(g)** displays the calculated band structure, along with the integrated density of states (DOS). In qualitative agreement with the experiments, the initial DOS maxima at the VB and CB onsets are followed by stronger maxima, out of which the occupied states are closer to the VB onset than the empty states to the CB onset. These two DOS maxima correspond to van Hove singularities at the $\Gamma$ point with contributions from the valence and conduction band (as opposed to the VB and CB onsets at the zone boundary) in combination with following bands. Focusing on the frontier states, the calculated wavefunctions and the associated STM image simulations (**Fig. 3(h,i)**) show an excellent agreement with the measurements and confirm our assignment.

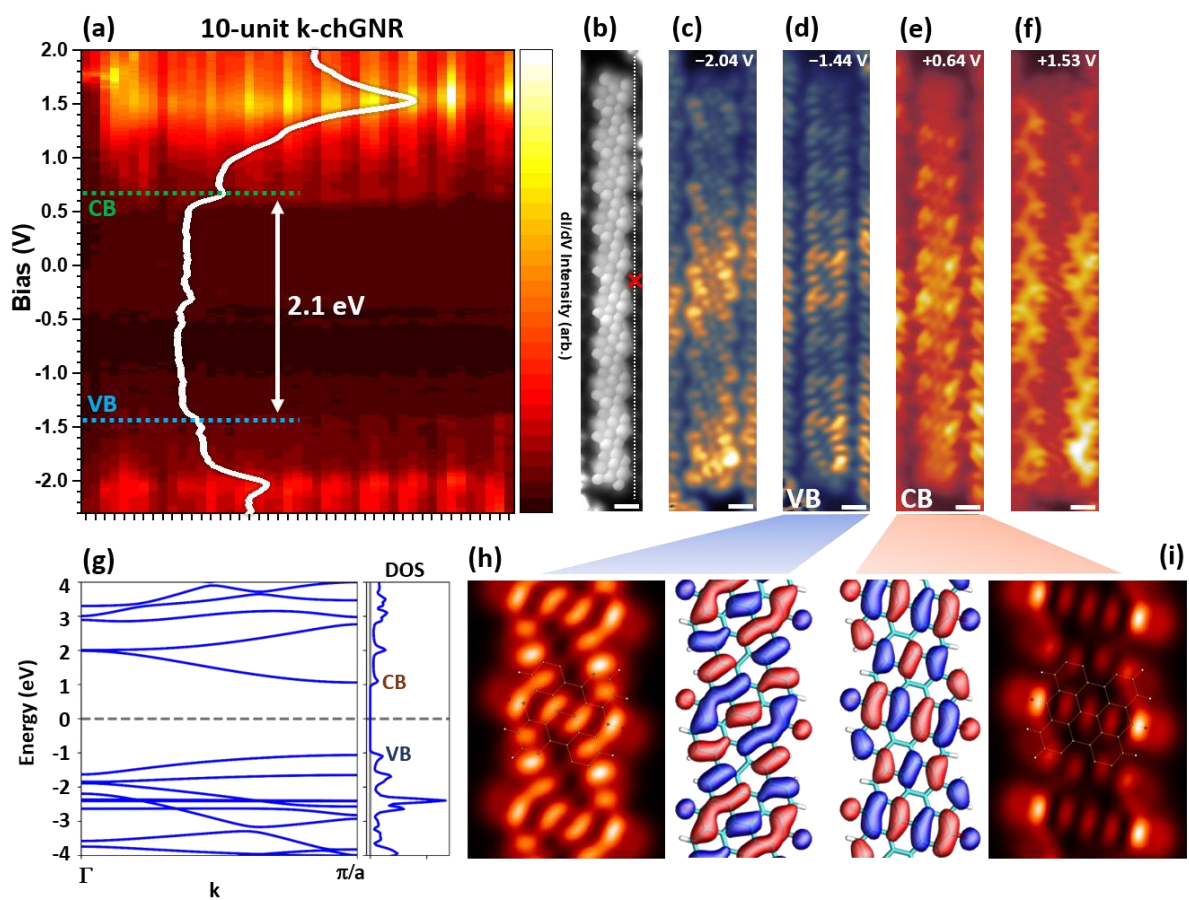

**Figure 3. Electronic properties of pre-oxidized protected GNRs**. (a) 45-point dI/dV line scan along a 10-unit ketone nanoribbon. The onsets of the valence and conduction bands are indicated with dashed lines, with a measured band gap of 2.1 eV. One of the spectra, recorded at the point marked with a red cross, is overlaid onto the line scan. (b) BR-STM image of the same 10-unit nanoribbon, with the dI/dV line scan position indicated. Constant height, CO tip, $V_{bias}$ = 5 mV. (c)-(f) Constant height dI/dV images of the four strongest resonances observed in the dI/dV line scan. All recorded with the same CO tip, all scale bars are 500 pm. (g) Band structure calculations for the ketone-functionalized ribbons (aligned with respect to the midgap energy), along with the integrated density of states. (h) Wavefunction and simulated STM image of the orbital at the valence band onset and (i) at the conduction band onset (the simulated images consider tips with 95 % $p_xp_y$ and 5 % $s$-wave character).

*Hydrogenation and annealing for deprotection*

As previously mentioned, the k-chGNRs are found to be particularly stable (in contrast to the pristine chGNRs[14]) due to the presence of carbonyl (C=O) groups that allow the formation of more Clar sextets without the concomitant creation of unpaired radicals. However, disrupting this system with atomic hydrogen may result in the formation of hydroxy derivatives that can in turn be involved in dehydration to afford the pristine chGNRs upon annealing. This was tested upon samples with both dominantly 'normal' and metal-coordinated k-chGNRs, with the latter being shown in **Fig. 4**.

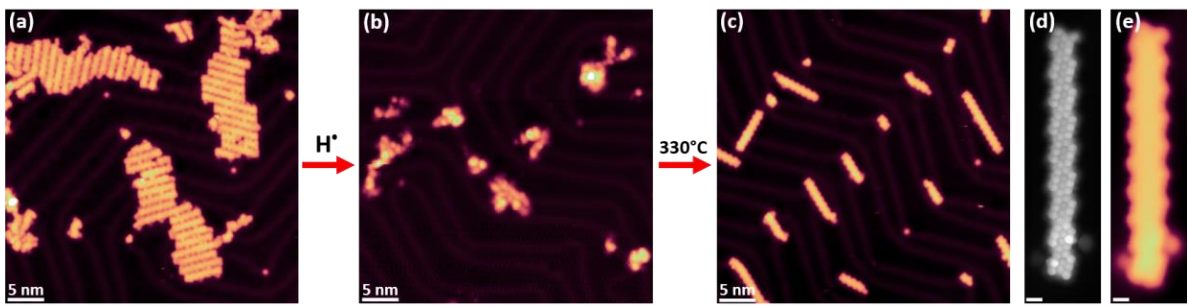

**Figure 4. STM analysis of GNRs along the various steps in the synthesis process of its pre-oxidized (protected) form, hydrogenation and annealing (deprotection).** (a) Overview STM image of mostly metal-organic ketone GNRs after their synthesis on Au(111). I = 100 pA, U = −0.1 V. (b) Overview STM image of the same GNR sample after exposure to atomic hydrogen. I = 50 pA, U = −1.0 V. (c) STM image of the same sample after annealing to 330 °C. The GNRs mostly consist of pristine sections, with some remaining ketone defects. I = 50 pA, U = −0.5 V. (d) BR-STM (Constant height, CO tip, 5 mV) and (e) STM (I = 100 pA, U = −0.5 V) of one of the NRs from the post-annealed sample. Two neighbouring metal-organic ketone defects are observed on the two sections at the lower end of the ribbon. Scale bars are both 500 pm.

Hydrogenation of the k-chGNRs (**Fig. 4(b)**) under similar conditions to the previous experiments (see methods section) results in a multitude of isolated structures, many of which are distorted into small "three-dimensional" objects with little resemblance to nanoribbons. However, annealing this sample to 330°C clearly demonstrates that the ribbons' backbones are not destroyed by the hydrogenation process, as elongated, straight structures are once again formed (**Fig. 4(c)**). Examination of these structures via BR-STM (**Fig. 4(d)**) reveals that they are, in fact, now mostly pristine GNRs, with a small number of ketone defects still remaining, most of which appear coordinated to Au atoms (e.g. the two units at the bottom of the GNR in **Fig. 4(d,e)**). These defects have the same appearance as those observed when exposing the hydrogenated pristine GNRs to air. Similar results are observed independently of the ratio of 'normal' to Au-coordinated ribbons. The high level of conversion (approx. 78%) may be further improved with tweaks to the hydrogenation process; however, this result is remarkable in and of itself. While the resulting GNRs in this case are relatively short in length, this is due not to the hydrogenation process itself, but to the short average length of the k-chGNRs that they began as. Improvements in the synthesis of the k-chGNRs, which may be presumably achieved with a reduced amount of *Z*-isomer as well as with optimized growth parameters, should also result in longer, higher quality pristine GNRs after the hydrogenation and annealing processes.

These results demonstrate that chemically protected GNRs can be easily converted back to pristine GNRs, and this method could be applicable to other protecting groups/functionalities that have different properties, as long as the central carbon at the zigzag edge is still present. Reducing the level of hydrogenation may also allow for statistical mixtures of functionalized/pristine GNRs that have different properties.

**Discussion**

It is important to note that hydrogenation cannot necessarily be used as a 'cure' for uncontrollably oxidized GNRs, since some of the common defects after exposing chGNRs to air involved modifications of the carbon-backbone structure and even the loss of carbon atoms that cannot be recovered through hydrogenation.[14] However, we have shown that this method allows the application of protection/deprotection strategies, as commonly utilized in conventional organic chemistry, to on-surface synthesis. That is, hydrogenation can be used both to protect chemically labile graphene nanoribbons, as well as to deprotect pre-oxidized nanoribbons that possess a significantly larger band gap and in-built resistance to oxidizing atmospheres, in either case allowing for a reconversion into pristine chiral graphene nanoribbons. Most importantly, this approach may be extrapolated to different graphene nanoribbons and carbon-based nanostructures, as well as to different functional groups that are involved in pre-protected forms, altogether bringing the exploitation of the unique characteristics of zigzag edges in carbon materials a step closer to scalable applications.

**Methods**

*Synthesis of (E/Z)-2,2'-dibromo-10H,10'H-[9,9'-bianthracenylidene]-10,10'-dione (k-DBBA)*

*General methods for solution synthesis*

All reactions were carried out under argon using oven-dried glassware. TLC was performed on Merck silica gel 60 $F_{254}$; chromatograms were visualized with UV light (254 and 360 nm). Flash column chromatography was performed on Merck silica gel 60 (ASTM 230-400 mesh). $^1$H NMR were recorded at 300 MHz (Varian Mercury 300). Low-resolution electron impact mass spectra were determined at 70 eV on a HP-5988A instrument. High-resolution mass spectra (HRMS) were obtained on a Micromass Autospec spectrometer. Commercial reagents were purchased from ABCR, GmbH, Aldrich Chemical Co., and were used without further purification. THF and $Et_2O$ was purified by a MBraun SPS -800 Solvent Purification System. Anthrone **6** was prepared following a published procedure shown in **fig. S10**.[37]

*Synthesis of (E/Z)-k-DBBA* (**fig. S11**)

Over a solution of anthrone **6** (500 mg, 1.84 mmol) in $CH_2Cl_2$ (20 mL), DBU (0.97 mL, 6.50 mmol) was added and the mixture was stirred for 10 min. Then, iodine (514 mg, 2.02 mmol) was added and the resulting solution was stirred in absence of light for 24 h. Then, aqueous HCl (10%, 10 mL) and saturated aqueous solution of $Na_2S_2O_3$ (20 mL) were added, the organic layer was separated and the aqueous phase was extracted with $CH_2Cl_2$ (2 x 20 mL). The combined organic extracts were dried over $Na_2SO_4$, filtered and evaporated under reduced pressure. The residue was purified by column chromatography ($SiO_2$; $CH_2Cl_2$), affording a mixture of (*E/Z*)-2,2'-dibromo-10H,10'H-[9,9'-bianthracenylidene]-10,10'-dione (k-DBBA) (1:1, 302 mg, 61%) as an orange solid (**fig. S12**).

Isomer 1: $^1$H NMR (300 MHz, CDCl$_3$) δ: 8.14 (d, *J* = 8.0 Hz, 1H), 8.00 (d, *J* = 8.6 Hz, 1H), 7.62 (dd, *J* = 8.3, 1.9 Hz, 1H), 7.55 – 7.43 (m, 1H), 7.31 – 7.25 (m, 1H), 7.20 (d, *J* = 1.9 Hz, 1H), 7.03 (t, *J* = 7.1 Hz, 1H) ppm.

Isomer 2: $^1$H NMR (300 MHz, CDCl$_3$) δ: 8.10 (d, *J* = 7.8 Hz, 1H), 7.97 (d, *J* = 9.3 Hz, 1H), 7.55 (dd, *J* = 8.3, 1.8 Hz, 1H), 7.50 – 7.37 (m, 1H), 7.23 – 7.13 (m, 1H), 7.15 (d, *J* = 1.8 Hz, 1H), 7.03 (t, *J* = 7.1 Hz, 1H) ppm.

MS (EI) m/z (%): 542 (M$^+$, 100), 381 (13), 353 (16), 324 (33), 162 (50). HRMS: $C_{28}H_{14}O_2Br_2$: calculated: 539.9361, found: 539.9368

*On-surface synthesis*

Low temperature STM measurements were performed using a commercial Scienta-Omicron LT-STM at 4.3 K. The system consists of a preparation chamber with a typical pressure in the low $10^{-10}$ mbar regime and an STM chamber with a pressure in the $10^{-11}$ mbar range. The Au(111) crystal was cleaned via cycles of argon sputtering and annealing (720 K). In order to form pristine (3,1)-chGNRs, a racemic mixture of chiral 2,2'-dibromo-9,9'-bianthracene (DBBA) precursors was deposited onto the Au(111) surface (held at room temperature) via sublimation at 433 K. Following this, the sample was typically annealed to approximately 600-620 K for 20 minutes in order to form the nanoribbons. To form ketone-chGNRs, a mixture of *E* and *Z* ketone-DBBA precursors (see supporting information for details on the synthesis of these precursors) was deposited onto the room temperature Au(111) crystal via sublimation at 448 K. In order to form flat ketone-chGNRs, the sample was typically heated to around 670-700 K after deposition. Heating straight to this temperature typically yielded more metal-organic GNRs than 'normal' GNRs. If the sample is instead heated to a moderately high temperature (e.g. 650 K) and then annealed in steps of 15-20 K up to 700 K, then more 'normal' GNRs are usually found on the surface. In our case, the sample was cooled down in between each step to investigate whether GNRs had been formed, but it is not clear whether this is necessary. More investigation would be required to understand the kinetics of the formation of the different types of k-chGNR.

### *GNR hydrogenation*

Hydrogenation of the samples was achieved with a simple hydrogen cracking source with a leak valve and a narrow inlet that flows through a tungsten tube into the chamber. The preparation chamber was first filled to a pressure of $1 \times 10^{-7}$ mbar, after which the tungsten tube was heated (via e-beam heating) to around 2800 K with a heating power of 80 W (corresponding to an acceleration voltage of 1000 V, 80 mA emission current) – this temperature is believed to be high enough to split the molecular hydrogen into atomic hydrogen, but there is no way of measuring the flux of atomic H on the simple hydrogen source used for these experiments. The sample was then placed in front of the source for 1-2 minutes in the experiments described in the main text. In both cases, the sample was then annealed to around 570-600 K to dehydrogenate the nanoribbons and form pristine chGNRs.

### *GNR exposure to ambient conditions*

For the air exposures, a sample of nanoribbons on Au(111) was transferred to the fast entry load-lock chamber, after which the chamber was vented to the air (laboratory room temperature and average relative air humidity in San Sebastián are around 21°C and 75%, respectively). The sample was then transferred back into the UHV chambers for STM analysis. To remove most of the contamination that comes along with the air exposure, the sample was annealed to 200°C before its new STM analysis.

### *STM characterization*

All STM measurements shown were performed at 4.3 K. The tunnelling current set points and bias voltages used are mentioned in the figure captions for each image. To obtain BR-STM images, the tip was functionalised with a CO molecule that was picked up from the Au(111) surface. To deposit CO, the STM chamber was filled to a pressure of approximately $1 \times 10^{-8}$ mbar of pure CO, and the Au(111) sample was exposed via opening the STM radiation shields. Each deposition was limited in time, such that the temperature of the sample did not exceed 7 K due to radiative heating. Picking up a CO molecule from the surface was achieved via scanning at high tunnelling current set points (1 nA) and negative bias voltages (typically −0.5 V to −1.0 V). A sudden change in apparent tip height and resolution was usually observed upon picking up a CO. BR-STM images were obtained by scanning a CO tip over the molecule at constant height with a low bias voltage (usually 5-8 mV). dI/dV measurements were recorded with the internal lock-in of the system. Typical oscillation parameters: 828 Hz, 20 mV.

*Theoretical calculations*

DFT calculations were performed using the FHI-AIMS code [38] using the hybrid exchange-correlation functional B3LYP [39] to describe the electronic properties of different GNR models in the gas phase. In all the calculations, we employed the light settings for the atomic basis sets. The atomic structures were relaxed until the total forces reached below $10^{-2}$ eV Å$^{-1}$. For the infinite systems with 1D periodic boundary condition (pristine and ketone-functionalized GNRs) a Monkhorst–Pack [40] grid of 18 × 1 × 1 was used to sample the Brillouin zone, and for the Au-coordinated ketone-functionalized ribbons with 2D periodicity, a grid of 18x18x1 k-points. All band structure calculations were carried out by 50 k-points. Theoretical dI/dV maps were calculated by the DFT Fireball [41] program package and Probe Particle Scanning Probe Microscopy (PP-STM) code [42,43] for a CO-like orbital tip, represented by 95% $p_xp_y$ and 5% $s$-like wave character.

27. Held, P. A., Fuchs, H. & Studer, A. Covalent-Bond Formation via On-Surface Chemistry. *Chem. Eur. J.* **23**, 5874–5892 (2017).

28. Song, S. *et al.* On-surface synthesis of graphene nanostructures with π-magnetism. *Chem. Soc. Rev.* **50**, 3238–3262 (2021).

29. Liu, J. & Feng, X. Synthetic Tailoring of Graphene Nanostructures with Zigzag-Edged Topologies: Progress and Perspectives. *Angew. Chem. Int. Ed.* **59**, 23386–23401 (2020).

30. Li, J. *et al.* Single spin localization and manipulation in graphene open-shell nanostructures. *Nat Commun* **10**, 200 (2019).

31. Mishra, S. *et al.* Topological frustration induces unconventional magnetism in a nanographene. *Nat. Nanotechnol.* **15**, 22–28 (2020).

32. Zuzak, R. *et al.* On-Surface Synthesis of Chlorinated Narrow Graphene Nanoribbon Organometallic Hybrids. *J. Phys. Chem. Lett.* **11**, 10290–10297 (2020).

33. Zuzak, R., Jančařík, A., Gourdon, A., Szymonski, M. & Godlewski, S. On-Surface Synthesis with Atomic Hydrogen. *ACS Nano* **14**, 13316–13323 (2020).

34. de Oteyza, D. G. *et al.* Substrate-Independent Growth of Atomically Precise Chiral Graphene Nanoribbons. *ACS Nano* **10**, 9000–9008 (2016).

35. Merino-Díez, N. *et al.* Transferring axial molecular chirality through a sequence of on-surface reactions. *Chem. Sci.* **11**, 5441–5446 (2020).

36. Merino-Díez, N. *et al.* Unraveling the electronic structure of narrow atomically-precise chiral graphene nanoribbons. *The Journal of Physical Chemistry Letters* **9**, 25–30 (2018).

37. Itoh, T., Matsuno, M., Kamiya, E., Hirai, K. & Tomioka, H. Preparation of Copper Ion Complexes of Sterically Congested Diaryldiazomethanes Having a Pyridine Ligand and Characterization of Their Photoproducts. *J. Am. Chem. Soc.* **127**, 7078–7093 (2005).

**Acknowledgments:**

**Funding:** Research was supported by Agencia Estatal de Investigación, grant nos. PID2019-107338RB-C62 and PID2019-107338RB-C63, by the European Union's Horizon 2020 program, grant nos. 863098 and 635919, by the Gobierno Vasco, grant no. PIBA_2020_1_0036, by the Xunta de Galicia (Centro Singular de Investigación de Galicia, 2019-2022, grant no. ED431G2019/03), the European Regional Development Fund, the Praemium Academie of the Academy of Science of the Czech Republic, GACR project no. 20-13692X and the Czech Nanolab Research Infrastructure supported by MEYS CR, project no. LM2018110.

**Author contributions:**

DP and DGO conceived the research. JCE, MVV and DP synthesized the reactants. JL, ABL, TW, MSGM and DGO performed on-surface synthesis, STM characterization and analysis. SE and PJ performed the theoretical calculations. All authors contributed to the scientific discussion, as well as to the review and editing of the manuscript.

Supplementary Information for

**Circumventing the Stability Problems of Graphene Nanoribbon Zigzag Edges**

James Lawrence[1,2,†], Alejandro Berdonces-Layunta[1,2,†], Shayan Edalatmanesh,[3] Jesús Castro-Esteban,[4] Tao Wang[1,2], Mohammed S. G. Mohammed[1,2], Manuel Vilas-Varela,[4] Pavel Jelinek,[3,*]

Diego Peña[4,*], Dimas G. de Oteyza[1,2,5,*]

[1]Donostia International Physics Center; San Sebastián, Spain

[2]Centro de Física de Materiales; San Sebastián, Spain

[3]Institute of Physics, Czech Academy of Sciences; Prague, Czech Republic

[4]Centro Singular de Investigación en Química Biolóxica e Materiais Moleculares (CiQUS) and Departamento de Química Orgánica, Universidade de Santiago de Compostela; Santiago de Compostela, Spain

[5]Ikerbasque, Basque Foundation for Science; Bilbao, Spain

* Corresponding author. Email: jelinekp@fzu.cz; diego.pena@usc.es; d_g_oteyza@ehu.es

† These authors contributed equally to this work

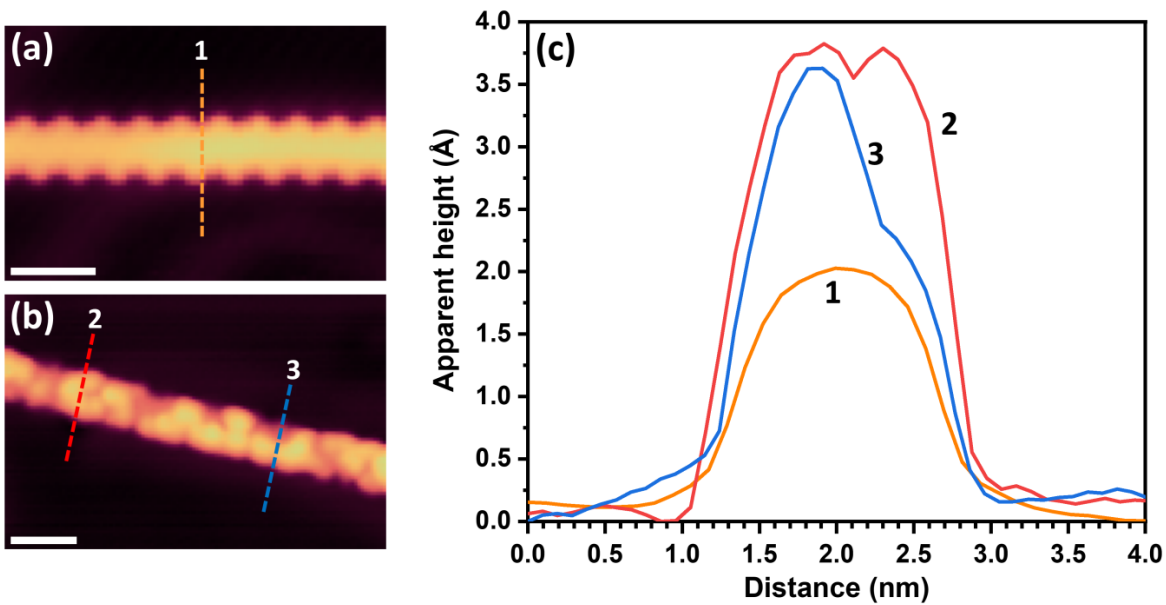

**Fig. S1.**
STM images of (a) pristine and (b) hydrogenated chiral nanoribbons ((3,1)-chGNRs). (a) I = 1.5 nA; U = −0.5 V. (b) I = 23 pA; U = −2.0 V. Scale bars are both 2 nm. (c) Relative apparent height profiles of the lines shown in (a) and (b). The hydrogenated nanoribbons appear to be approximately twice as high as the pristine nanoribbons; the darker sections are a similar height to the pristine GNRs, which suggests that they are not hydrogenated to the same extent.
24

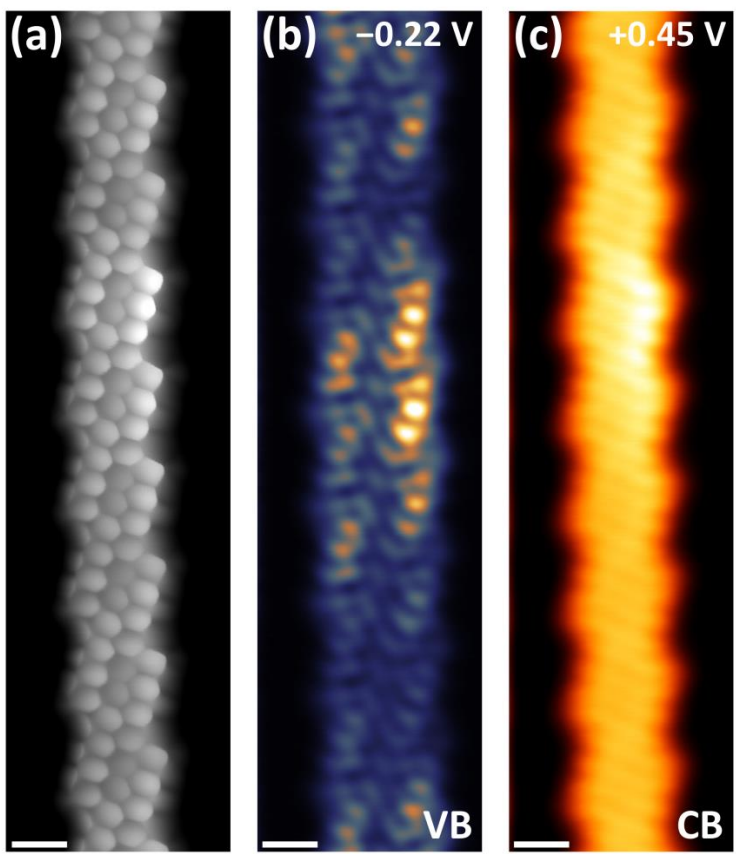

**Fig. S2.**
(a) BR-STM image of a pristine GNR section after hydrogenation, air exposure and post-annealing to 300°C. Constant height, CO tip, $V_{bias}$ = 5 mV. (b) and (c) Constant height dI/dV images (CO tip) of the same pristine GNR section, demonstrating that the features of the valence and conduction bands are preserved after these treatments. The brighter regions of the images are due to the underlying herringbone reconstruction of the Au(111).

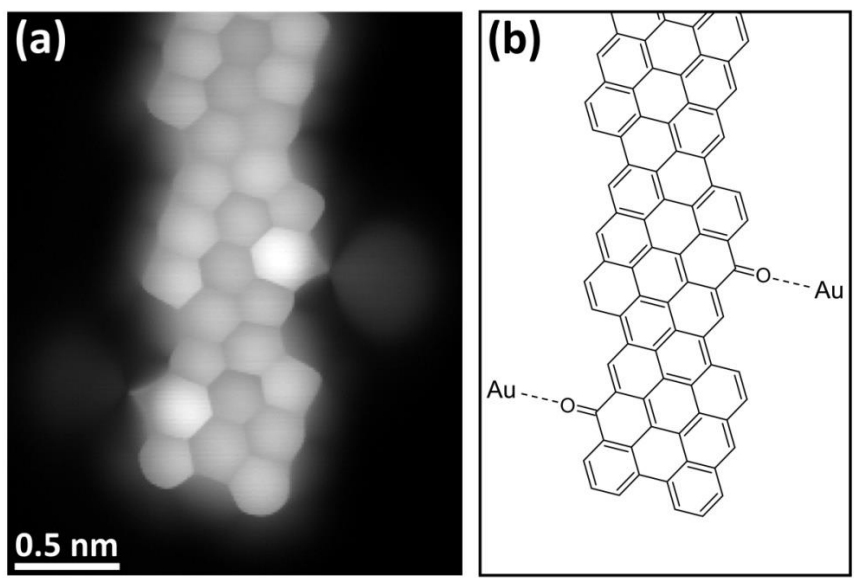

**Fig. S3.**
(a) BR-STM image and (b) chemical model of a pair of ketone-metal defects at one end of a (mostly) pristine chiral nanoribbon. Constant height, CO tip, $V_{bias}$ = 5 mV. These defects are found in various different configurations (alone, in pairs on the same GNR section, etc.). This example is from the ketone GNR sample that was hydrogenated and then annealed. This implies that not all of the oxygens are removed from the GNRs during the hydrogenation/annealing process. The same defects were often observed with the pristine GNR sample that was hydrogenated, exposed to air and annealed. It follows that these defects probably occur on positions that were not hydrogenated, and thus had reacted with oxygen during the air exposure.

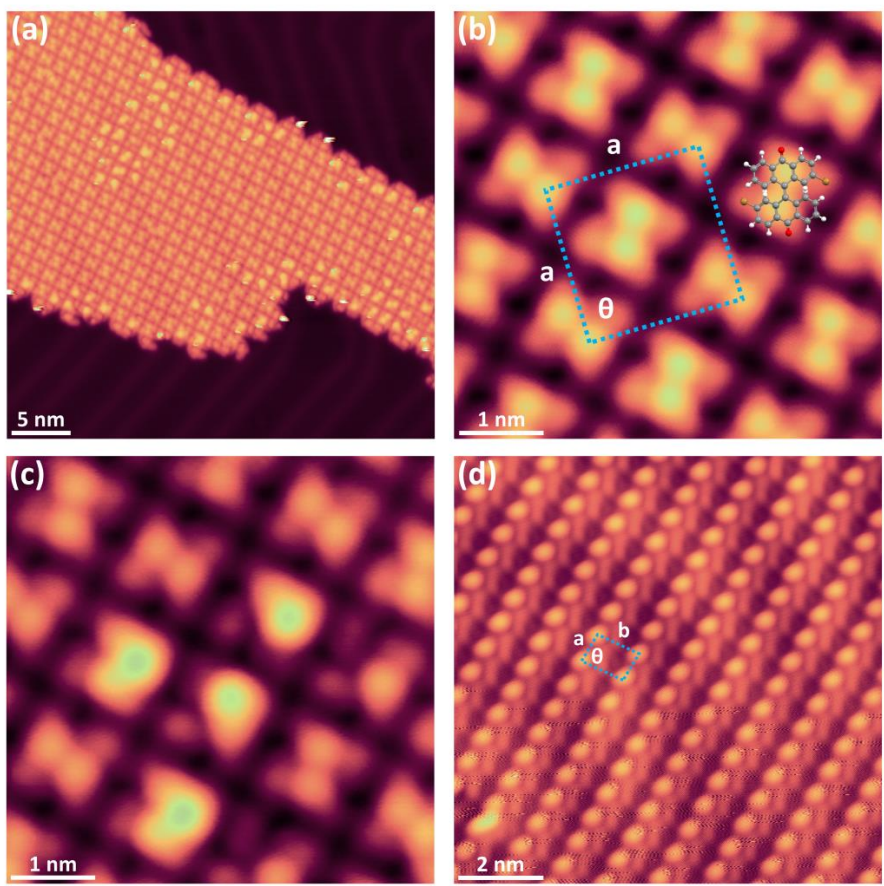

**Fig. S4.**
(a) STM image of an island of ketone-GNR precursors at a relatively low molecular coverage. (b) Zoom of the same island, showing the regular structure and its unit cell. **a** = 1.87 ± 0.03 nm. θ = 90 ± 1°. A scaled gas-phase optimised model of a trans precursor molecule is overlaid in one of its possible conformations for reference. (c) Zoom of a region of an island with 'defects' – possibly different conformations of molecules or a mixture of enantiomers. (d) STM image of a packing that was typically observed at higher coverages. **a** = 0.67 ± 0.02 nm; **b** = 1.06 ± 0.01 nm; θ = 85 ± 2°. All four STM images in this figure were recorded with the following parameters: I = 50 pA; $V_{bias}$ = −0.5 V.

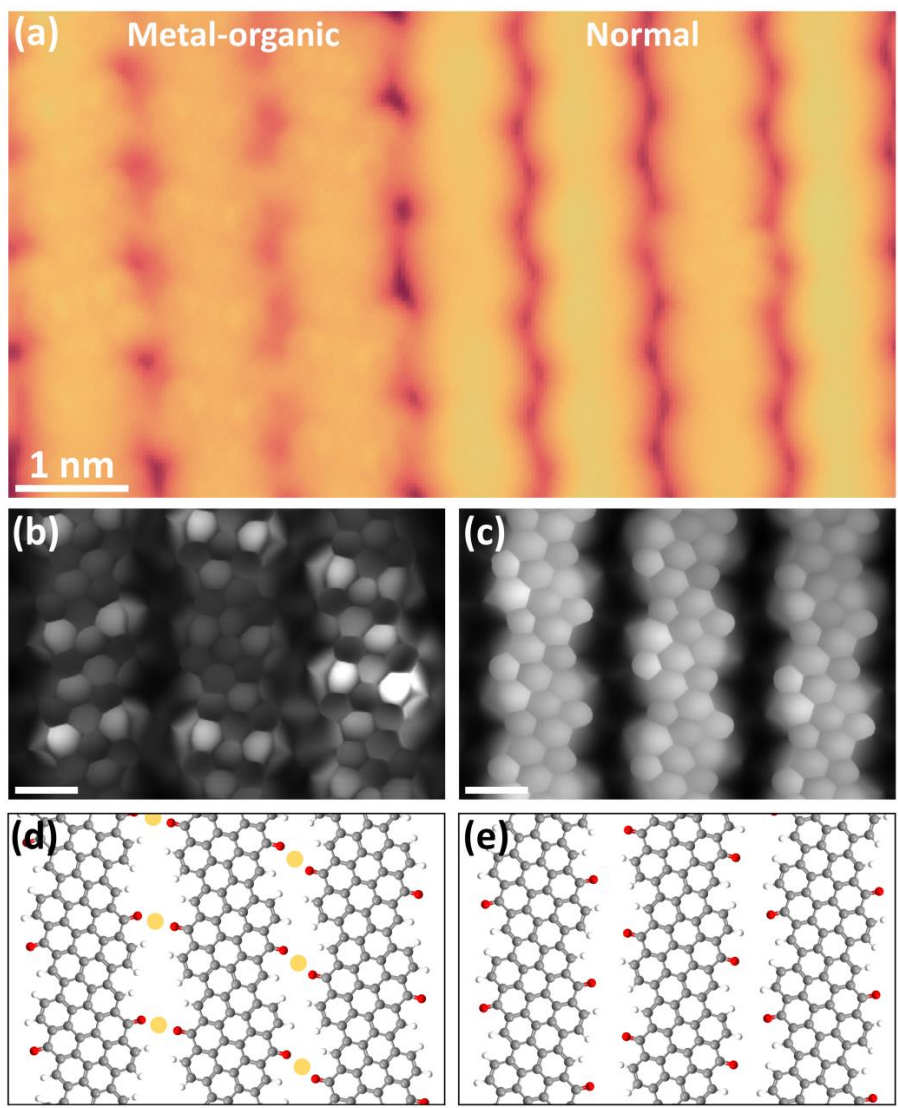

**Fig. S5.**

Comparison between the structures of the two types of ketone-GNRs that were observed. (a) Typical STM image of a self-assembled island that contains both types of ketone-GNR. I = 50 pA, $V_{bias}$ = −0.5 V. (b) BR-STM image of a metal-organic and (c) normal section of a ketone-NR island. Both constant height, CO tip, $V_{bias}$ = 4 mV and 5 mV, respectively. Scale bars are both 500 pm. (d) and (e) Models of the metal-organic and normal islands, with suggested positions for gold adatoms in the metal-organic model.

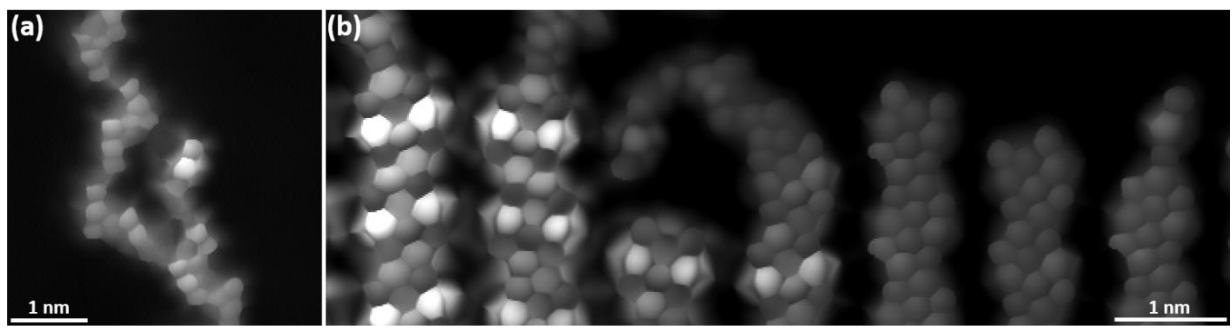

**Fig. S6.**
(a) BR-STM image (Constant height, CO tip, 5 mV) of a covalently bonded/fused cluster of anthrone-related fragments. (b) BR-STM image (Constant height, CO tip, 5 mV) of the termini of several ketone-GNRs, showing how they are often found fused with an anthrone-related fragment.

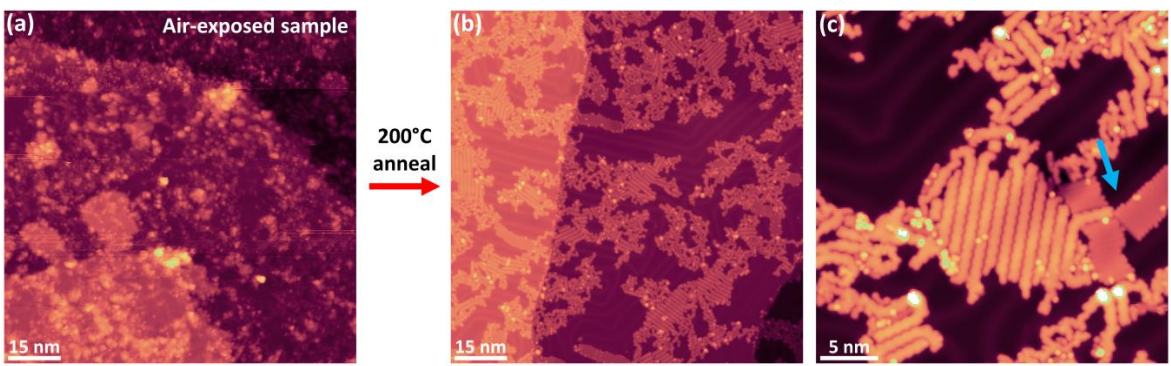

**Fig. S7.**
(a) Overview STM image of a ketone-GNR sample that was exposed to air for 24 minutes, without any further treatment. I = 50 pA; $V_{bias}$ = 1.5 V. (b) and (c): STM images of the same sample after annealing to 200°C for 1 hour. (b) I = 20 pA; $V_{bias}$ = −0.8 V. (c) I = 30 pA; $V_{bias}$ = −0.8 V. Although most have desorbed, a significant number of contaminants are still present on the surface after the annealing treatment. Islands of alkyl chain contamination from the ambient conditions are indicated in (c) with a blue arrow.

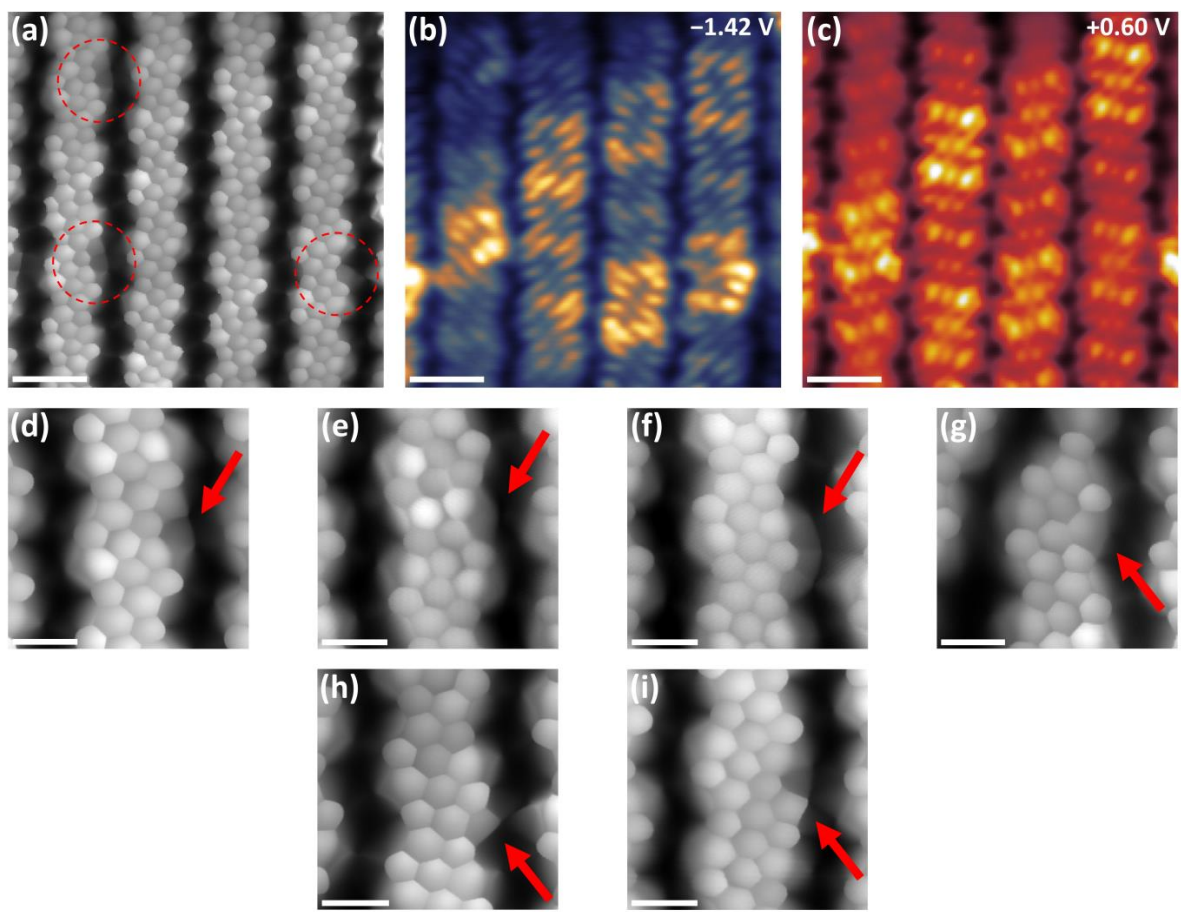

**Fig. S8.**

(a) BR-STM image (Constant height, CO tip, $V_{bias}$ = 5 mV) of an island of ketone-GNRs that were exposed to the atmosphere (24 mins) and post-annealed to 200°C for 1 hour. Defects are highlighted with dashed red circles. (b) and (c) Constant height dI/dV images of the same area (CO tip) at the approximate energies of the valence and conduction bands, respectively. While there are clear changes at the position of the defects, the general spatial distribution of the states is maintained. The appearance of the conduction band onset is relatively unaffected by the defects. Scale bars in (a) – (c) are 1 nm. (d) – (i) BR-STM images (Constant height, CO tip, $V_{bias}$ = 5-8 mV) of various defects that are found after exposing the ketone-GNRs to ambient conditions (plus post-annealing). Many oxidation defects may be found in positions that already possessed defects prior to air exposure, e.g. sections in which ketone groups were missing. This would cause them to be more reactive and readily oxidise. Scale bars in (d) – (i) are 0.5 nm.

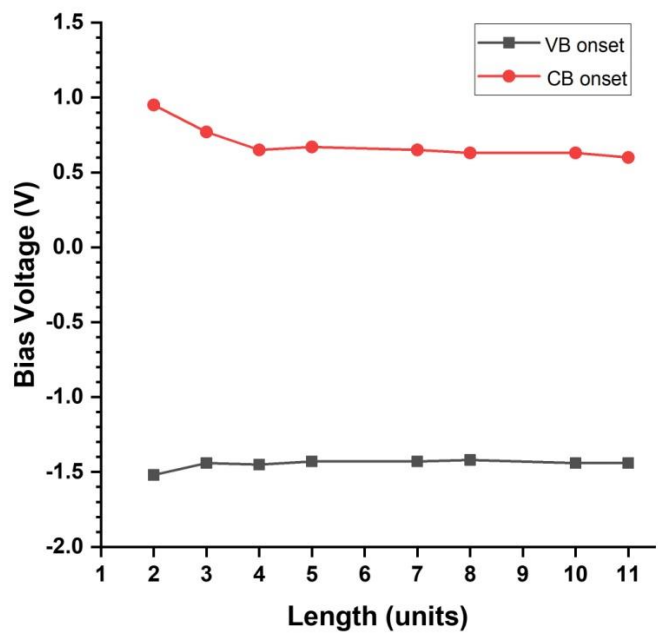

**Fig. S9.**
Dependence of the VB and CB onset positions on the length of the ketone-GNRs. The band gap very quickly saturates after 4 or 5 units in length to a value between 2.04 eV and 2.10 eV, with the smallest gap recorded for the 11-mer.

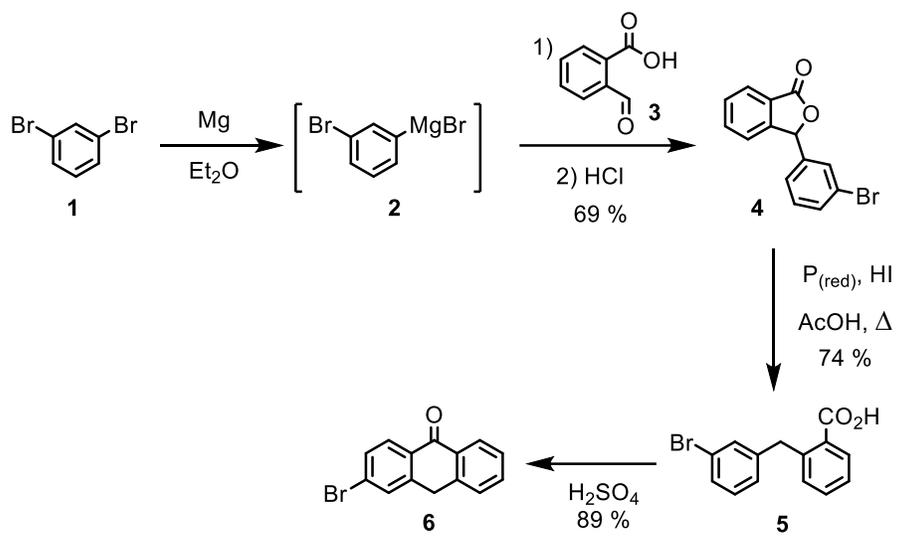

**Fig. S10.**
Synthesis of anthrone **6**.

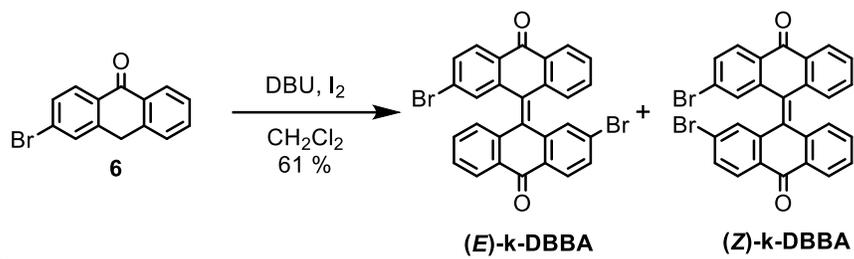

**Fig. S11.**
Synthesis of (E/Z)-k-DBBA

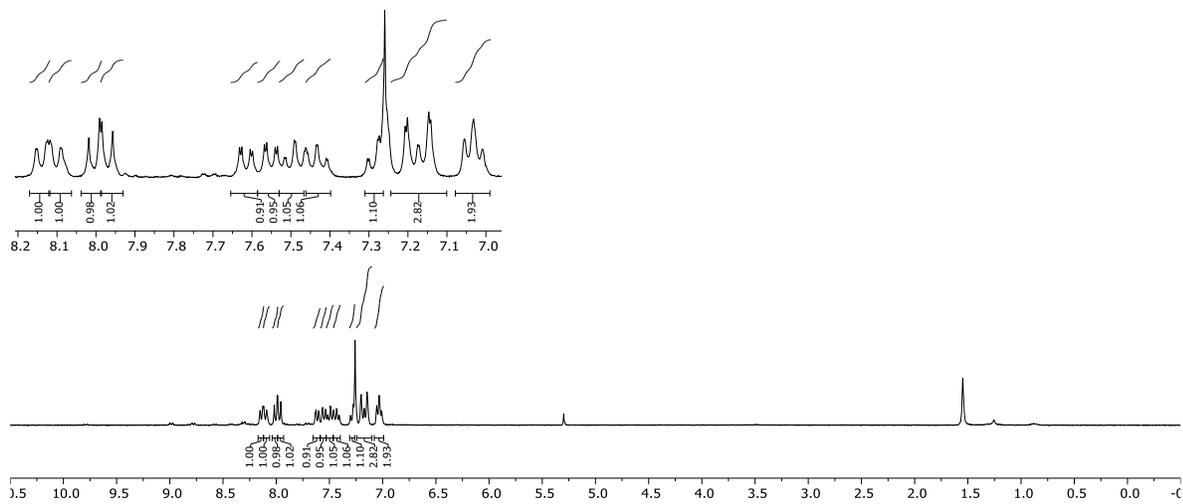

**Fig. S12.**
¹H NMR spectra of (E/Z)-k-DBBA